\documentclass[12pt]{article}

\usepackage{longtable}
\usepackage{pifont}
\usepackage{amsmath}
\usepackage{mathtools}
\usepackage{caption}
\usepackage{subcaption}
\usepackage{graphicx}
\usepackage{placeins} 
\usepackage{verbatim}
\usepackage{xr}
\usepackage{booktabs}
\usepackage{multirow}
\usepackage{amssymb}
\usepackage{lmodern}
\usepackage{rotating}
\usepackage{float}
\usepackage[T1]{fontenc}
\usepackage[a4paper,left=2.5cm,right=2.5cm,top=3cm,bottom=3cm]{geometry}
\usepackage[flushleft]{threeparttable}
\usepackage{authblk}

\usepackage[style=authoryear-comp, backend=bibtex, maxcitenames=2, maxbibnames=99 ,dashed=false]{biblatex}
\bibliography{Literatur}

\usepackage{hyperref}

\begin{document}

\title{Bias through time-varying covariates in the analysis of cohort stepped wedge trials: a simulation study}
\author[1]{Jale Basten}
\author[2]{Katja Ickstadt}
\author[1]{Nina Timmesfeld}
\affil[1]{Ruhr-University of Bochum, Department of Medical Informatics, Biometry and Epidemiology}
\affil[2]{TU Dortmund University, Faculty of Statistics}
\date{}

\maketitle

\begin{abstract}
In stepped wedge cluster randomized trials (SW-CRTs), observations collected under the control condition are, on average, from an earlier time than observations collected under the intervention condition. In a cohort design, participants are followed up throughout the study, so correlations between measurements within a participant are dependent of the timing in which the observations are made. Therefore, changes in participants’ characteristics over time must be taken into account when estimating intervention effects. For example, participants’ age progresses, which may impact the outcome over the study period. Motivated by an SW-CRT of a geriatric care intervention to improve quality of life, we conducted a simulation study to compare model formulations analysing data from an SW-CRT under different scenarios in which time was related to the covariates and the outcome. The aim was to find a model specification that produces reliable estimates of the intervention effect. Six linear mixed effects (LME) models with different specification of fixed effects were fitted. Across 1000 simulations per parameter combination, we computed mean and standard error of the estimated intervention effects. We found that LME models with fixed categorical time effects additional to the fixed intervention effect and two random effects used to account for clustering (within-cluster correlation) and multiple measurements on participants (within-individual correlation) seem to produce unbiased estimates of the intervention effect even if time-varying confounders or their functional influence on outcome were unknown or unmeasured and if secular time trends occurred. Therefore, including (time-varying) covariates describing the study cohort seems to be avoidable.
\end{abstract}

\section{Introduction} \label{chap::introduction}

In cluster randomized trials (CRTs), the unit of randomization is the cluster, rather than the participants themselves \parencite{Murray1998}. They are frequently designed to evaluate the effect of an intervention administered at the cluster level, such as hospitals, practices or geographical units, to minimize treatment contamination \parencite{Torgerson2001}, or for other logistic, financial or ethical reasons.
\par
Clustering must be accounted for in the design to avoid an under-powered study and in the analysis to avoid under-estimated standard errors and inflated type I error for the intervention effect \parencite{Murray1998, Turner2017_PART2, Eldridge2012}. This is due to the fact that observations from participants in the same cluster are usually correlated, resulting in higher sample sizes of CRTs compared to the sample size in an individually randomized trial \parencite{Hooper2016}.
\par
\begin{figure}[!htb]
 \centerline{\includegraphics[width=250pt,height=10pc]{"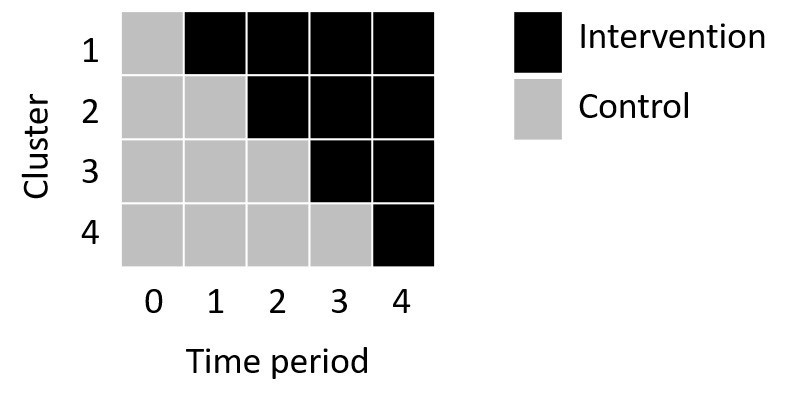"}}
 \caption{Schematic representation of a standard SWD with four intervention sequences (steps), where one cluster per step changes from control to intervention.\label{fig:SWD}}
\end{figure}

However, longitudinal CRTs such as the stepped wedge design (SWD; Figure~\ref{fig:SWD}) can lead to a reduction of sample size \parencite{Woertman2013,Hooper2016,Matthews2017}. The key feature of stepped wedge cluster randomized trials (SW-CRTs) is the unidirectional crossover of clusters from the control to intervention conditions on a staggered schedule. The first time point usually corresponds to a baseline measurement where none of the clusters receive the intervention of interest \parencite{Hussey2007}. In contrast to the cluster randomized parallel design, the SWD randomly assigns the time of transition to the intervention rather than the group membership of the cluster. Thus, the SWD allows for comparisons within clusters as well as between clusters (horizontal and vertical comparisons). Vertical approaches compare outcomes of clusters in the intervention condition with the outcomes of clusters in the control condition within the same period. As the order of rollout is randomized, each of these comparisons is randomized. Horizontal approaches compare outcomes before and after crossover to the intervention condition. These are non-randomized comparisons that are confounded with time period \parencite{Thompson2017_B, Hargreaves2015}. Therefore, time effects are a well-known problem of SWDs \parencite{Hemming2015_2}.

Depending on the intervention and patient recruitment and based on the terminology of Feldman and McKinlay\parencite{Feldman1994}, different approaches can be chosen for the SWD: On the one hand, in a cross-sectional design, participants contribute one measurement during the study, i.e., in every time step different participants will be considered; on the other hand, in closed cohort designs all participants are identified at the onset of the trial and participate from start till end. In addition, Copas et al \parencite{Copas2015} discussed a third option, the open cohort design, where a substantial number of participants are identified and participate from the beginning, but some may leave during the trial and others may become eligible and be exposed for some time. In a medical context, for example, the cohort design usually includes chronically ill patients who correspond to the target group of the intervention over several time periods, while a cross-sectional design observes newly ill patients who are no longer part of the target population soon after the intervention took place. The methodological focus of SW-CRTs has been on cross-sectional designs, which was originally proposed by Hussey and Hughes \parencite{Hussey2007}. But in practice, a majority of SW-CRTs are studies of closed or open cohort designs \parencite{Martin2016}.
\par
In an SW-CRT, observations collected under the control condition are, on average, from an earlier calendar time than observations collected under the intervention condition. Changes external to the trial may create underlying secular trends, e.g., public policies, seasonal fluctuations or COVID-19-related lockdowns. Simulation studies have shown that such external time effects need to be taken into account to avoid highly biased estimates of the intervention effect and type I and II errors \parencite{Thompson2017_B, Nickless2018, Rennert2021}. According to the CONSORT (Consolidated Standards of Reporting Trials) extension for SW-CRTs \parencite{Hemming2019}, time effects represent one of the most potential confounders of SW-CRTs leading to bias \parencite{Barker2016}, and thus require special attention both at the design and analysis stage. The guideline indicates that the analysis of SW-CRTs should adjust for external time effects irrespective of their statistical significance; failure to do so risks biasing the estimate of the intervention effect, which could lead to declaring an intervention effective when it is ineffective or ineffective when it is effective. It is therefore essential to report if and how time effects were allowed for \parencite{Hemming2019}.
\par
In an SW-CRT that follows a cross-sectional design, we need to distinguish between two types of correlation between measurements: the within-period correlation, i.e., the correlation between two within-cluster observations collected during the same period, and the between-period correlation, i.e., the correlation between two within-cluster observations collected in different time periods \parencite{Li2020_Overview}. The latter play a role in horizontal comparison and can be influenced by external time effects. In cohort SW-CRTs, in addition to the two correlations between two within-cluster observations, correlations between two repeated measurements from the same individual occur. We refer to these as within-individual correlation \parencite{Li2020_Overview}. In addition to external time trends, the within-individual correlation is influenced by changes in participants' characteristics over time. For example, participants' age progresses and they may experience certain life changes that impact the outcome over the study period. We refer to this in the following as internal time effects. Because cohort SW-CRTs require further consideration of internal time effects and are more complicated to analyze due to the more complex correlation structure than SW-CRTs in cross-sectional designs, there should be clarity in reporting trials as cohort or cross-sectional designs \parencite{Martin2016}.
\par
Given a lack of guidance for trialists on the correct approach to evaluate cohort SW-CRTs and to deal with internal time effects, we investigate the impact of external and internal time effects in estimating the participant specific effect of the intervention, i.e., the average intervention effect for a randomly selected participant in a randomly selected cluster in SW-CRT with closed and open cohort data. This effect can be estimated with linear mixed effects (LME) regression via (restricted) maximum likelihood estimation \parencite{Hussey2007, Turner2017_PART2}. The intervention effect is conditional on the random effects used to account for the correlation of outcomes of participants who belonged to the same cluster (within-cluster correlation) and repeated measurements on the same participant over time (within-individual correlation). The purpose of this simulation study is to find a model specification that produces unbiased estimates of participant specific intervention effects in closed and open cohort SW-CRTs.

\section{Motivating examples}\label{chap::motivation}

Changes of the cohort over time - the magnitude of which in turn depends on the study duration and additionally on the attrition rate in the case of the open cohort design - lead to the fact that time-shifted control and intervention do not take place under the same conditions. Thus, the observations under control conditions may not be comparable to the observations under intervention conditions not only because of external time effects, but also because of internal time effects. In an open cohort design, for example, younger and healthier patients might be included in the cohort, while older and sicker patients might no longer be able to meet the requirements for study participation or might have died during the study.
\par
In the cohort design, study participants are usually chronically ill patients who will not recover completely and thus remain part of the target population of the intervention over the study period. For this reason, frail older adults are often considered in cohort SW-CRTs. An example is the ``frail older Adults: Care in Transition'' (ACT) trial (The Netherlands National Trial Register, NTR2160) \parencite{Muntinga2012, Hoogendijk2015}.
\par
The cohort SW-CRT investigates the effectiveness of a geriatric care model on the quality of life of community-dwelling frail older adults and ran over a 24-month period. Quality of life was measured by the 12-item Short Form questionnaire (SF-12) which is divided in two domains: a mental health component score (MCS) and a physical health component score (PCS). In the ACT trial, only the MCS was used and an increase over time was observed \parencite{Twisk2016}. The increase in MCS can be due to different causes or also to combinations of them:
\begin{enumerate}
  \item the intervention increases the MCS and the proportion of observations under intervention condition increases in the SW-CRT,
  \item secular trends lead to a change in mental health independent of the intervention, and
  \item the study cohort changes over time or by the attrition of participants which impacts the MCS independent of the intervention.
\end{enumerate}
All three causes might have an impact on the MCS in the same or in different directions. For example, the intervention might be effective and leads to an improvement in mental health. But the increase could also be explained by a general increase over the time of the MCS independent of the intervention, e.g., due to generally higher levels of prosperity in the society, more leisure activities for the elderly, and much more. In contrast, as people age over the study period, they could become more frail or may experience certain life changes which impact their mental health negatively, such as coping with a serious illness or losing a loved one. Thus, progression of age in the study cohort might have a negative impact on mental health while the intervention and secular trends have a positive impact. However, it could also be possible that the observed cohort becomes younger and healthier over the study period as frail and older participants die and/or younger and healthier participants join the study. To estimate the intervention effect without bias, the intervention effect would have to be adjusted for external and internal time effects (cause 2 and 3).
\par
In the ACT trial, an increase of the outcome variable over time irrespective of the intervention was observed in the whole population \parencite{Twisk2016}. Because at the end of the trial, there are more patients in the intervention condition, this increase over time is wrongly allocated to the intervention when time is not taken into account. For this reason, time has been explicitly included in the LME model. Furthermore, time since the start of the intervention was included. To this end, the study group constructed a categorical variable indicating the time since the start of the intervention (6, 12, 18, and 24 months), with all control measurements as a reference category \parencite{Hoogendijk2015}. For the ACT trial, Twisk et al \parencite{Twisk2016} conclude that there is neither a short-term nor a long-term intervention effect, but the observed differences between the stepped wedge arm on average over time are due to differences at baseline, when all participants in the four allocation groups (group 1 starts intervention 6 months after baseline, group 2 another 6 months later, and so on) are still under control conditions. In the result paper of the effectiveness evaluation published in 2015, all models calculated were adjusted for age, sex and baseline variables on which the allocation groups differed at baseline (educational level, region and frailty index score) \parencite{Hoogendijk2015}. Regarding internal time effects due to changes in the cohort over time, insight into changes in the frailty status of participants during the study period would have been interesting. However, the authors admitted that they unfortunately did not measure frailty during follow-up but only at baseline. Furthermore, up to 18\% of subjects dropped out from one follow-up measurement to the next. This loss to follow-up will probably not be missing at random, since, for example, no further follow-up measurements can be collected from participants who died. Thus, because of loss to follow-up, it is also likely that the cohort considered at baseline, in which a large proportion was under control conditions, cannot be directly compared with the cohort toward the end of the study, in which a large proportion is under intervention conditions.
\par
Our simulation study aims to investigate which model specification yields unbiased intervention effect estimates in spite of external and internal time effects.

\section{Methods}\label{chap::methods}

\subsection{Simulation study methods}

According to the motivating example, we again consider the SF-12 as outcomes to investigate the impact of a physical activity intervention on quality of life. In contrast to the motivating example, we consider the PCS instead of the MCS as outcomes. It ranges from 0 to 100 points, with 0 points representing the greatest possible physical health limitation and 100 points representing no physical health limitation.
\par
In the simulation study, $48$ clusters (= practices) with $8$ participants (= patients) each were randomized into $J = 4, 6,$ or $8$ groups. At each step $I/J$ clusters switch from the control to the intervention period. This study design corresponds to a standard SWD. That means, that in the first period all clusters are under control conditions and at each subsequent period $12$, $8$, or $6$ clusters start the intervention, respectively. Therefore, patients were followed for $5, 7, \text{ or } 9$ time periods, where one time period corresponds to a length of one half year.
\par
We use the subscript $i$ to denote practices where $i = 1,..., I$, $j$ to denote time steps (calendar time) where $j = 0,..., J$, and $k$ to denote patients in cluster $i$ where $k = 1,..., K_i$. In a closed cohort design, the clusters are of equal size, i.e., $K_i = 8$ $\forall$ $i$ $\in$ $\{1,...,I\}$. In an open cohort design, however, $K_i$ is a random number depending on the attrition rate of each cluster $i$. The notation $y_{ijk}$ represents the outcome (= SF-12 PCS) for patient $k$ at time step $j$ in practice $i$, and $x_{ij}$ is a binary indicator for whether the practice $i$ is in the intervention at time step $j$. A summary of mathematical notation is provided in Table~\ref{tab::notation}.

\begin{table}[!htb]
\centering
\caption{Summary of mathematical notation.}
\label{tab::notation}
\begin{tabular}{p{1cm}p{14cm}}
\toprule
$i$ & Cluster subscript $i = 1,..., I$\\ \hline
$j$ & Time steps (calendar time) $j=0,..., J$\\ \hline
$k$ & Participant subscript $k = 1,..., K_i$\\ \hline
$y_{ijk}$ & Outcome (= SF-12 PCS) for participant $k$ at time step $j$ in cluster $i$\\ \hline
$x_{ij}$ & Binary indicator for whether cluster $i$ is under the intervention condition at time step $j$\\ \hline
$\mu$ & Intercept of the LME model\\ \hline
$\theta$ & Intervention effect (coefficient of $x_{ij}$ in the LME model)\\ \hline
$c_i$ & Cluster specific random intercept with mean zero and variance $\sigma^2_c$\\ \hline
$d_{ik}$ & Participant specific random intercept with mean zero and variance $\sigma^2_d$\\ \hline
$e_{ijk}$ & Random error with mean zero and variance $\sigma^2_e$ conditional on  $c_i + d_{ik}$\\ \hline
$\beta_j$ & Coefficient of the binary indicator for categorical time $j$ in the LME model\\ \hline
$\beta_{age}$ & Coefficient of the participants' age in the LME model\\ \hline
$a_{ijk}$ & Age of participant $k$ at time step $j$ in cluster $i$\\ \hline
$\beta_{wid}$ & Coefficient of being widowed in the LME model\\ \hline
$w_{ijk}$ & Binary indicator for whether participant $k$ at time step $j$ in cluster $i$ is widowed\\ \hline
$l$ & Length of a step, i.e., time elapsing from one changeover point into intervention to the next\\
\bottomrule
\end{tabular}
\end{table}
\FloatBarrier

The intercept is constantly set to 70 in all data sets. However, we allow the number of steps $J$ to vary between $4, 6$ and $8$, and the size of the intervention effect to vary between $0, 0.25, 0.5, 0.75$ and $1$. A summary of the data scenarios simulated is given in Table~\ref{tab::parameter}.

\begin{table}[!htb]
\centering
\caption{Summary of simulation study data scenarios.}
\label{tab::parameter}
\begin{tabular}{p{6cm}p{9cm}}
\toprule
\textbf{Parameter} &\textbf{Value and description} \\
\midrule
Number of clusters ($= I$) & 48\\ \hline
Number of groups ($= J$) & 4, 6, 8. At each step $I/J$ clusters switch from the control to the intervention period. That means, if the number of groups is $J = 4, 6$ or $8$, then $12, 8$ or $6$ clusters are in one group, respectively.\\ \hline
Number of time periods ($= J+1$) & 5, 7, 9. The study duration is always one period longer than the number of groups, since in the first period (baseline period) all clusters are still observed under control conditions.\\ \hline
Cluster size ($= K_i$) & 8 participants in each cluster and time period in closed cohort design; random number in open cohort design. 
\\ \hline
Length of time period ($= l$) & one half year ($= 0.5$)\\ \hline
Intercept ($= \mu$) & 70\\ \hline
\multicolumn{2}{p{15cm}}{\textbf{Fixed Effects Parameterisations}}\\ \hline
Intervention effect ($= \theta$) & 0, 0.25, 0.5, 0.75, 1 common to all clusters \\ \hline
Participants' age ($= a_{ijk}$) & for every participant $k$ in cluster $i$ a random number of the uniform distribution with minimum equal to 18 and maximum equal to 102 for the baseline period $j=0$; for the further periods, the age continues to progress according to the period length.\\ \hline
Age effect ($= \beta_{age}$) & -0.25\\ \hline
Widowed status ($= w_{ijk}$) & probability for every person quarter $5\%$, if once widowed, participant remains widowed over the entire study duration.\\ \hline
Effect for being widowed ($= \beta_{wid}$) & -5 \\ \hline
External time effect, e.g., due to COVID-19-related lockdowns ($=\mathbf{\beta}= (\beta_0,..., \beta_J)^T$) & $(0, 0, -4, -4, -4, -5, -5, -5, -6)^T$; allows a different calendar time effect at each time step $j$.\\ \hline
\multicolumn{2}{p{15cm}}{\textbf{Random Effects Parameterisations}}\\ \hline
Between-cluster variance ($= \sigma_c^2$) & 10\\ \hline
Between-participant variance ($= \sigma^2_d$) & 10 \\ \hline
Random error variance ($= \sigma^2_e$) & 20\\ \hline
Attrition rate & 15\% (in case of open cohort data)\\ \hline
Age of joining participants & for every joining participant $k$ during the course of the study a random number of the uniform distribution with minimum equal to 18 and maximum equal to 96 for the individual baseline period; for further periods, the age continues to progress according to the period length.\\
\bottomrule
\end{tabular}
\end{table}
\FloatBarrier

In addition to the intervention, we assume that three other independent factors influence the outcome:
\begin{enumerate}
  \item[(1)] the patients' age as a continuous variable, which has a negative effect on SF-12 PCS
  \item[(2)] an indicator indicating whether the patient $k$ at time step $j$ in practice $i$ is widowed as a binary variable, also negatively affecting SF-12 PCS, and
  \item[(3)] a secular trend as a binary indicator for categorical time $j$, e.g., caused by COVID-19-related lockdowns.
\end{enumerate}

The participants' age at baseline ($j = 0$) is uniformly distributed with a range from 18 to 102, resulting in an average of 60 years at baseline. Therefore, the intervention provided patients of the practices physical activity opportunities for all ages from 18 to 102. As we assume that a period corresponds to one half year, each patient in the cohort becomes a half year older from one period to the next. Since the control periods are from an earlier calendar time than intervention periods, the participants are on average older under intervention than under control conditions. Thus, the outcome between the control and intervention periods cannot be compared without taking age progression into account. Since age has a negative effect on SF-12 PCS the estimate for the intervention effect will else be underestimated.
\par
The binary indicator for whether a participant is widowed at baseline occurs with a probability of 5\%. In the following steps, a further 5\% becomes widowed, whereby an already widowed participant remains widowed over the entire study duration. We assume a negative effect on outcome due to the listlessness in the mourning phase. Since the proportion of widowed persons increases during study, the effect has a negative impact on the SF-12 PCS, and since intervention takes place after control, this confounder also leads to an underestimation of the intervention effect in a direct comparison of control and intervention measurements.
\par
Also secular trends might impact the outcome over the study period. Such external time effects might be triggered, e.g., by influenza waves, seasonal fluctuations, or COVID-19-related lockdowns. Studies findings suggested lockdowns instituted during the COVID-19 pandemic may have had substantial adverse public health effects \parencite{Wilke2021, Chen2020}. Individuals’ access to sport and physical activity was hampered due to COVID-19-related lockdown restrictions. In many countries participation in community sport was cancelled during lockdowns. An Australian study \parencite{Eime2022} showed that the absence of playing competitive sport and training with friends, teams and within clubs severely impacted males and younger adults in particular. Sports clubs provide an important setting for individuals’ health and wellbeing which is why clubs require the capacity to deliver sport and individuals may need to regain the motivation to return. For our simulation study, we assume that a COVID-19-related lockdown at the beginning of period $j = 2$ has a negative impact on the SF-12 PCS. The negative effect increases slightly during pandemic ($(\beta_0,..., \beta_J)^T = (0, 0, -4, -4, -4, -5, -5, -5, -6)^T$).
\par
In practice, it is more realistic that a cohort is open and not closed. Patients can no longer be followed for various reasons and others are joined because, for example, they now meet the inclusion criteria. While the age and thus the proportion of widowed patients definitely increases when considering a closed cohort, this is not necessarily the case when considering an open cohort. Here, the cohort changes depending on which patients leave or join. An open cohort dataset is generated by replacing step-by-step 15\% of the participants with participants who now meet the inclusion criteria of the study cohort (attrition rate). We simulate that the oldest 15\% per cluster and period leave the cohort and younger patients join on average. The age of the new participants is uniformly distributed with a range from 18 to 96 years, resulting in an average of 57 years at baseline.

\subsection{Analysis models} \label{chap:AnalysisModels}

The time-varying cohort characteristics lead to the fact that time-shifted control and intervention do not take place under the same conditions. By computing LME models with different fixed effects, we examine how an unbiased estimate of the participant specific intervention effect can nevertheless be achieved for closed and open cohort data in SW-CRTs.
\par
For fitting LME models to data, via restricted maximum likelihood (REML), we use the R package \texttt{lmerTest} function \texttt{lmer}. In addition to computing the model (using \texttt{lme4::lmer}), it provides p-values and summary tables for LME models via Satterthwaite’s degrees of freedom method (approximating degrees of freedom for the t and F tests) \parencite{Kuznetsova2017}.
\par
The simplest analysis model used to analyze cohort data from an SW-CRT is an LME model with two random intercepts that account for clustering and for the induced correlation between the measurements obtained by the same individual, and adjusting for intervention effect as a fixed categorical variable \parencite{Baio2015}. In this case, the general framework for the simulation based
approach can be described as follows. Individual variability in the observed outcome data $y_{ijk}$ is described using a suitable distribution depending on the nature of the outcome and characterised by cluster specific, participant specific, and time specific mean $\mu_{ijk}$ and a variance $\sigma^2_e$ conditional on the random cluster and individual effect. The mean of the outcome is described by a linear predictor, on a suitable scale:
\begin{align}\label{AnalysisModel_WoAdj}
g(\mu_{ijk}) = \mu + c_i + d_{ik} + \theta x_{ij}.
\end{align}
As the outcome is continuous and normally distributed, the function $g(\cdot)$ is just the identity.
\par
Continuous and binary time-varying covariates, such as age or widowhood status, can be included in the LME model as fixed effects. Here, one could include the respective baseline value of the participant (model formulation~\ref{AnalysisModel_BaselineAdj}) or also adjust for the time-varying covariates by including the covariate value at each step (model formulation~\ref{AnalysisModel_StepAdj}). The mean value of the outcome is described by a linear predictor as follows:
\begin{align}
  g(\mu_{ijk}) &= \mu + c_i + d_{ik} + \theta x_{ij} + \beta_{age} a_{i0k} + \beta_{wid} w_{i0k}\label{AnalysisModel_BaselineAdj} \text{ or }\\
  g(\mu_{ijk}) &= \mu + c_i + d_{ik} + \theta x_{ij} + \beta_{age} a_{ijk} + \beta_{wid} w_{ijk}\label{AnalysisModel_StepAdj},
\end{align}
where $\beta_{age}$ is the coefficient of the participants' age, $a_{ijk}$ is the age of participant $k$ at time step $j$ in cluster $i$, $\beta_{wid}$ is the coefficient of being widowed, and $w_{ijk}$ is a binary indicator for whether participant $k$ at time step $j$ in cluster $i$ is widowed. Here, $j = 0$ indicates the baseline period in which all participants are observed under control conditions.
\par
To adjust for changes over time, a fixed effect $\beta_j$ corresponding to time period $j$ ($j \in 0,..., J, \beta_0 = 0$ for identifiability) could be included in model~\ref{AnalysisModel_WoAdj}:
\begin{align}\label{AnalysisModel_PeriodEff}
  g(\mu_{ijk}) = \mu + c_i + d_{ik} + \theta x_{ij} + \beta_{j}.
\end{align}
\par
The predictor in model formula~\ref{AnalysisModel_PeriodEff} could be extended including covariates describing the characteristic of the analysis cohort at baseline (model formulation~\ref{AnalysisModel_PeriodEffBaselineAdj}) or at each step (model formulation~\ref{AnalysisModel_PeriodEffStepAdj}):
\begin{align}
  g(\mu_{ijk}) &= \mu + c_i + d_{ik} + \theta x_{ij} + \beta_j + \beta_{age} a_{i0k} + \beta_{wid} w_{i0k}\label{AnalysisModel_PeriodEffBaselineAdj} \text{ or }\\
  g(\mu_{ijk}) &= \mu + c_i + d_{ik} + \theta x_{ij} + \beta_j + \beta_{age} a_{ijk} + \beta_{wid} w_{ijk}\label{AnalysisModel_PeriodEffStepAdj}.
\end{align}
\par
In Table~\ref{Tab:MixedModels} is an overview of the LME model formulations we compared.

\begin{table}[!htb]
\caption{Summary of analysis models.}
\label{Tab:MixedModels}
\centering
\begin{tabular}{p{0.15cm}p{6cm}p{8.85cm}}
    \toprule
    & \textbf{Description of adjustment} &\textbf{LME model formulations} \\
    \midrule
    1 & Without adjustment & $g$\tnote{$\dagger$}$(\mu_{ijk}) = \mu + c_i + d_{ik} + \theta x_{ij}$\\ \hline
    2 & Baseline adjustment & $g$\tnote{$\dagger$}$(\mu_{ijk}) = \mu + c_i + d_{ik} + \theta x_{ij} + \beta_{age} a_{i0k} + \beta_{wid} w_{i0k}$\\ \hline
    3 & Step-by-step adjustment & $g$\tnote{$\dagger$}$(\mu_{ijk}) = \mu + c_i + d_{ik} + \theta x_{ij} + \beta_{age} a_{ijk} + \beta_{wid} w_{ijk}$\\ \hline
    4 & Fixed categorical time effects & $g$\tnote{$\dagger$}$(\mu_{ijk}) = \mu + c_i + d_{ik} + \theta x_{ij} + \beta_{j}$\\ \hline
    5 & Fixed categorical time effects and baseline adjustment & $g$\tnote{$\dagger$}$(\mu_{ijk}) = \mu + c_i + d_{ik} + \theta x_{ij} + \beta_{j} +  \beta_{age} a_{i0k}  + \beta_{wid} w_{i0k}$\\ \hline
    6 & Fixed categorical time effects and step-by-step adjustment & $g$\tnote{$\dagger$}$(\mu_{ijk}) = \mu + c_i + d_{ik} + \theta x_{ij} + \beta_{j} +  \beta_{age} a_{ijk}  + \beta_{wid} w_{ijk}$\\
    \bottomrule
\end{tabular}
\begin{tablenotes}
\item[$\dagger$] As the outcome is continuous and normally distributed, the function $g(\cdot)$ is just the identity.
\end{tablenotes}
\end{table}
\FloatBarrier

\subsection{Estimands and performance measures}

We ran 1000 simulations for each combination of parameters. Simulations were run in R version 4.1.2. For every data set and analysis model, we collected the estimated fixed effects and their standard errors. Our most important performance measure is the bias of $\hat{\theta}$, i.e., the difference between the average intervention effect estimate over all simulations $\hat{\theta}$ and the true value $\theta$.
\par
The power can be calculated based on the proportion of significant simulations to all simulations. The basic principle underlying all simulation-based power analysis solutions can therefore be broken down into the following steps: (1) simulate new data sets, (2) analyze each data set and test for statistical significance, and (3) calculate the proportion of significance to all simulations \parencite{Kumle2021,Johnson2015,Thomas1996}. For an intervention effect of $\theta$ equal to zero, it can be checked whether the type I error is met.

\section{Results}\label{chap::results}

Since the primary goal of the cohort SW-CRT is to estimate the participant specific intervention effect, the estimate of interest will be the intervention effect parameter $\theta$ in an LME model. In the following three figures are the means and standard errors of 1000 intervention effect estimates for each combination of parameters shown. We varied the size of the true intervention effect $\theta = 0, 0.25, 0.5, 0 .75$, and $1$, as well as the number of steps $J = 4, 6$, and $8$ and thus the study duration of $J+1 = 5, 7$, and $9$ half years. The horizontal black line represents the true intervention effect in order to quickly identify whether the intervention effect is estimated without bias.
\par
\begin{figure}[!htb]\captionsetup[subfigure]{font=normalsize}
     \begin{subfigure}[t]{0.49\textwidth}
         \raisebox{-\height}{\includegraphics[width=\textwidth]{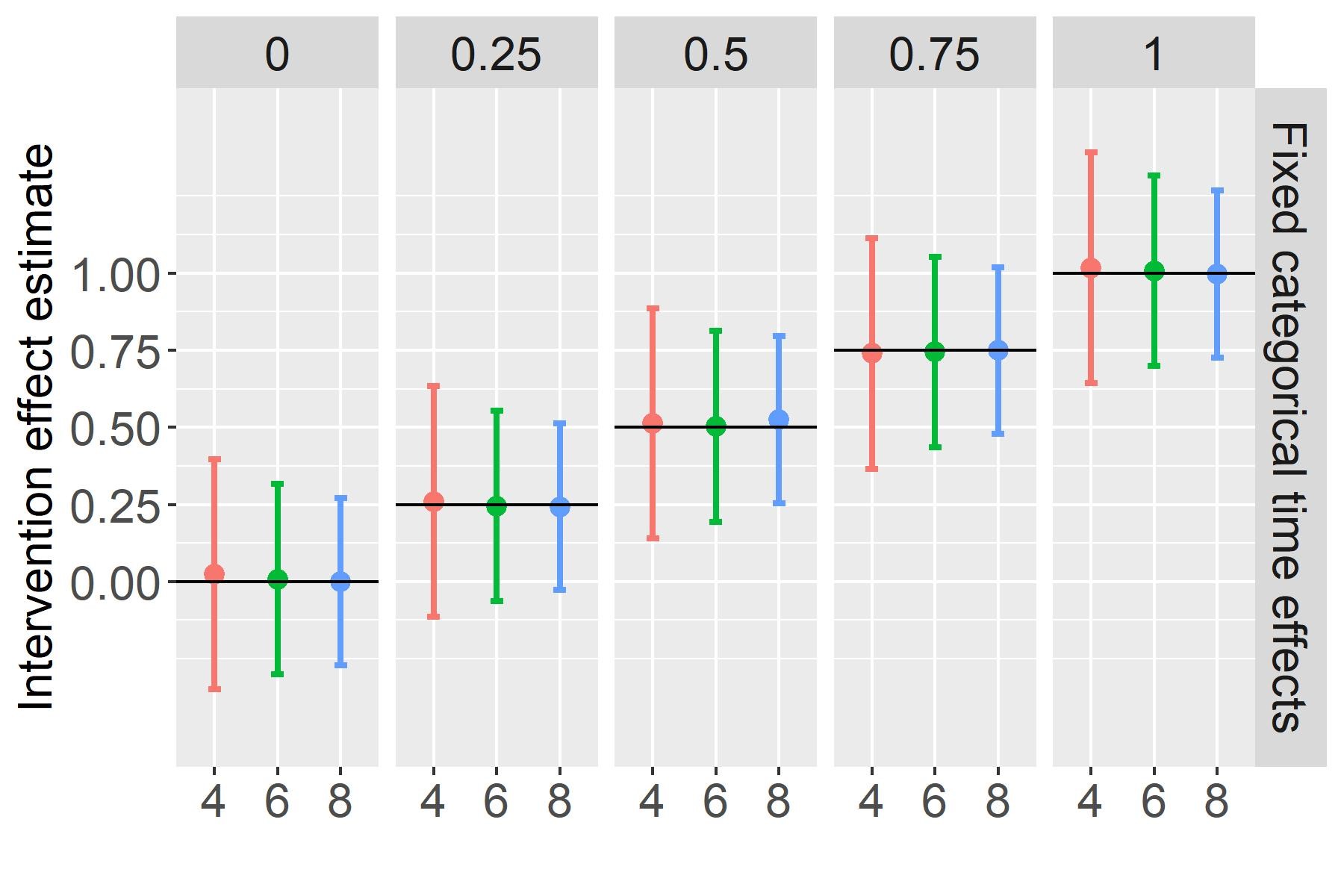}}
         \caption{Closed cohort data - Linear influence of continuous covariate}
         \label{fig:Estimate_linear_age_binary}
     \end{subfigure}
     \hfill
     \begin{subfigure}[t]{0.49\textwidth}
         \raisebox{-\height}{\includegraphics[width=\textwidth]{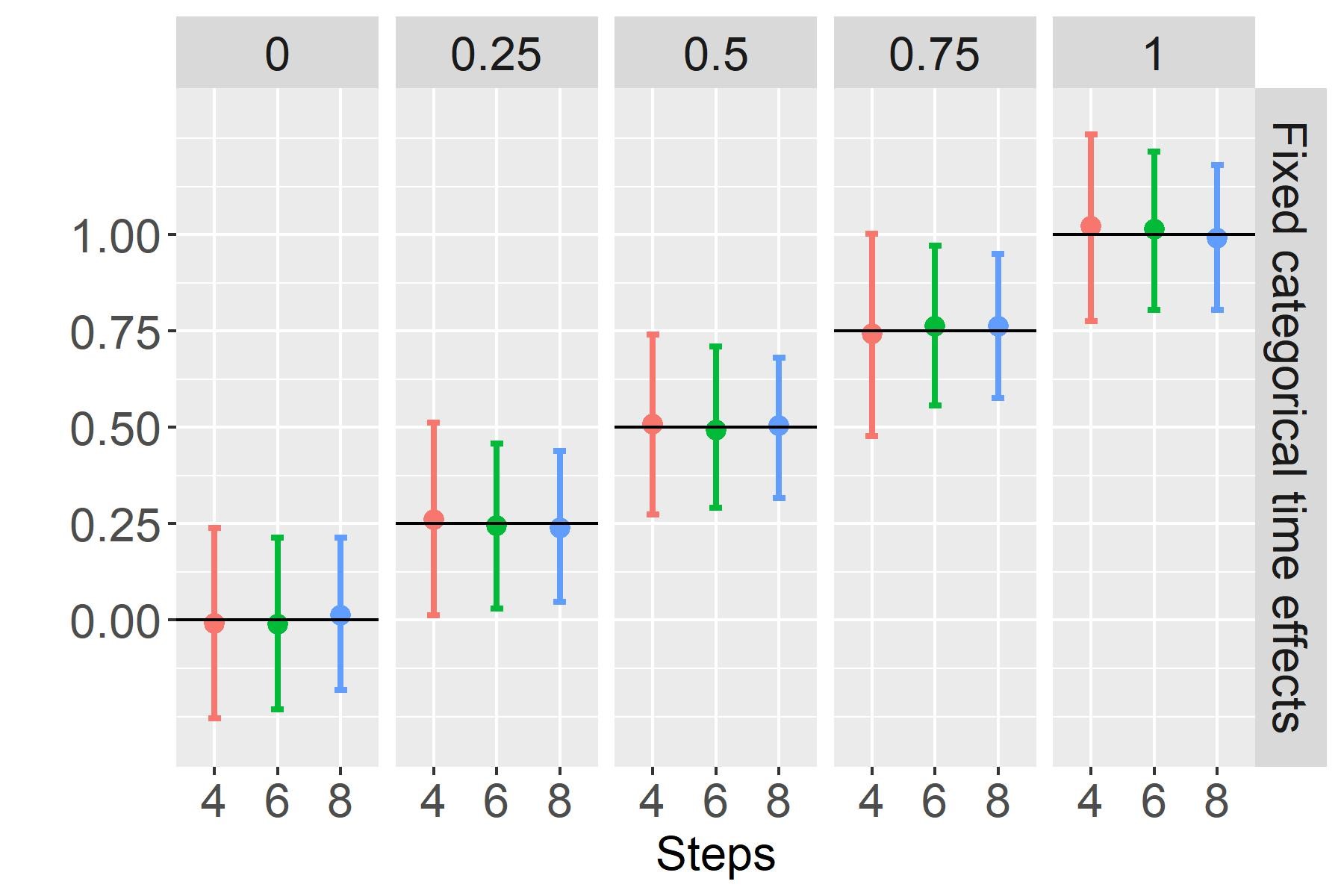}}
         \caption{Closed cohort data - Nonlinear influence of continuous covariate}
         \label{fig:Estimate_nonlinear_age_binary}
     \end{subfigure}.
     \begin{subfigure}[t]{0.49\textwidth}
         \centering
         \raisebox{-\height}{\includegraphics[width=\textwidth]{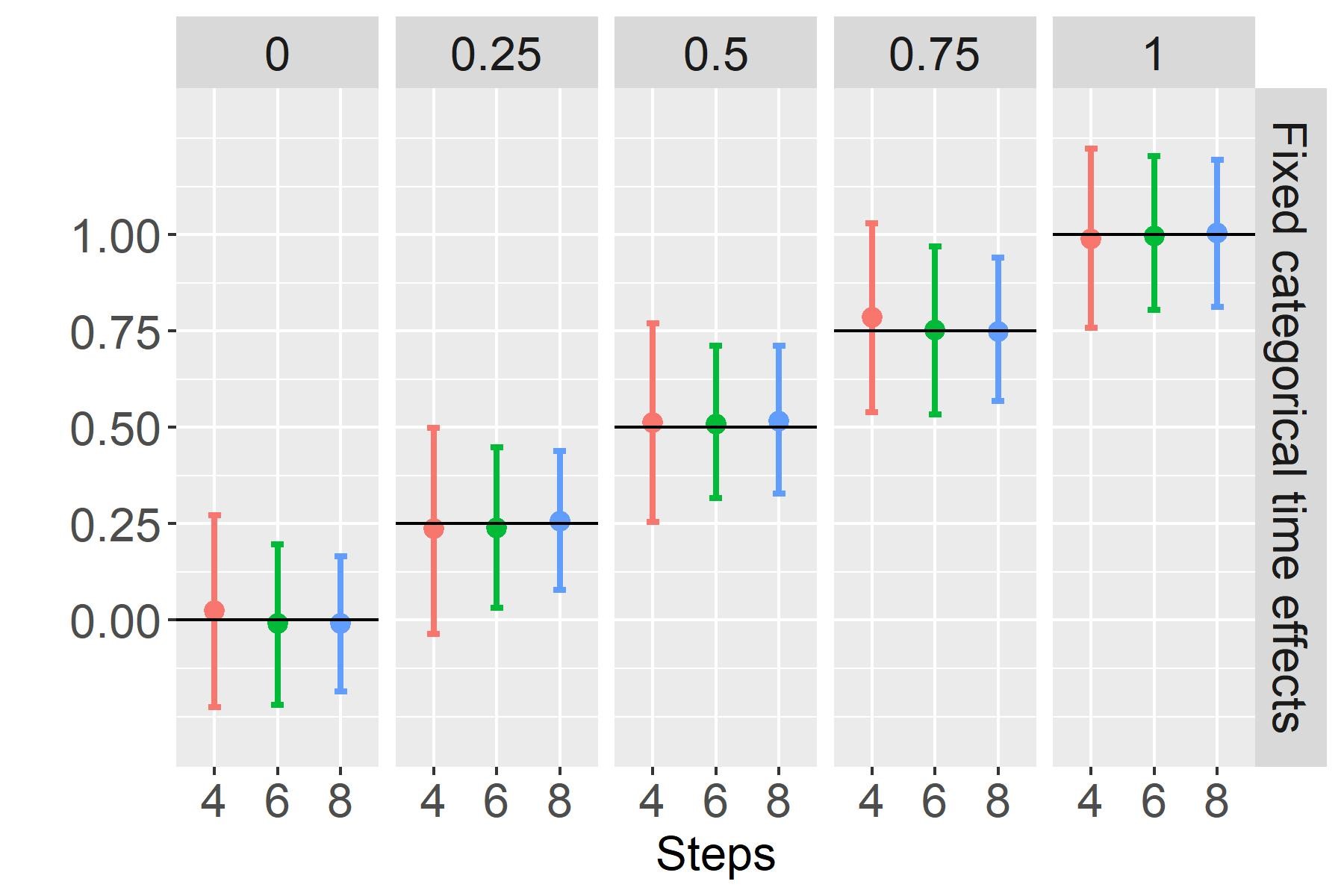}}
         \caption{Closed cohort data - Secular trend}
         \label{fig:Estimate_linear}
     \end{subfigure}
     \hfill
     \begin{subfigure}[t]{0.49\textwidth}
         \centering
         \raisebox{-\height}{\includegraphics[width=\textwidth]{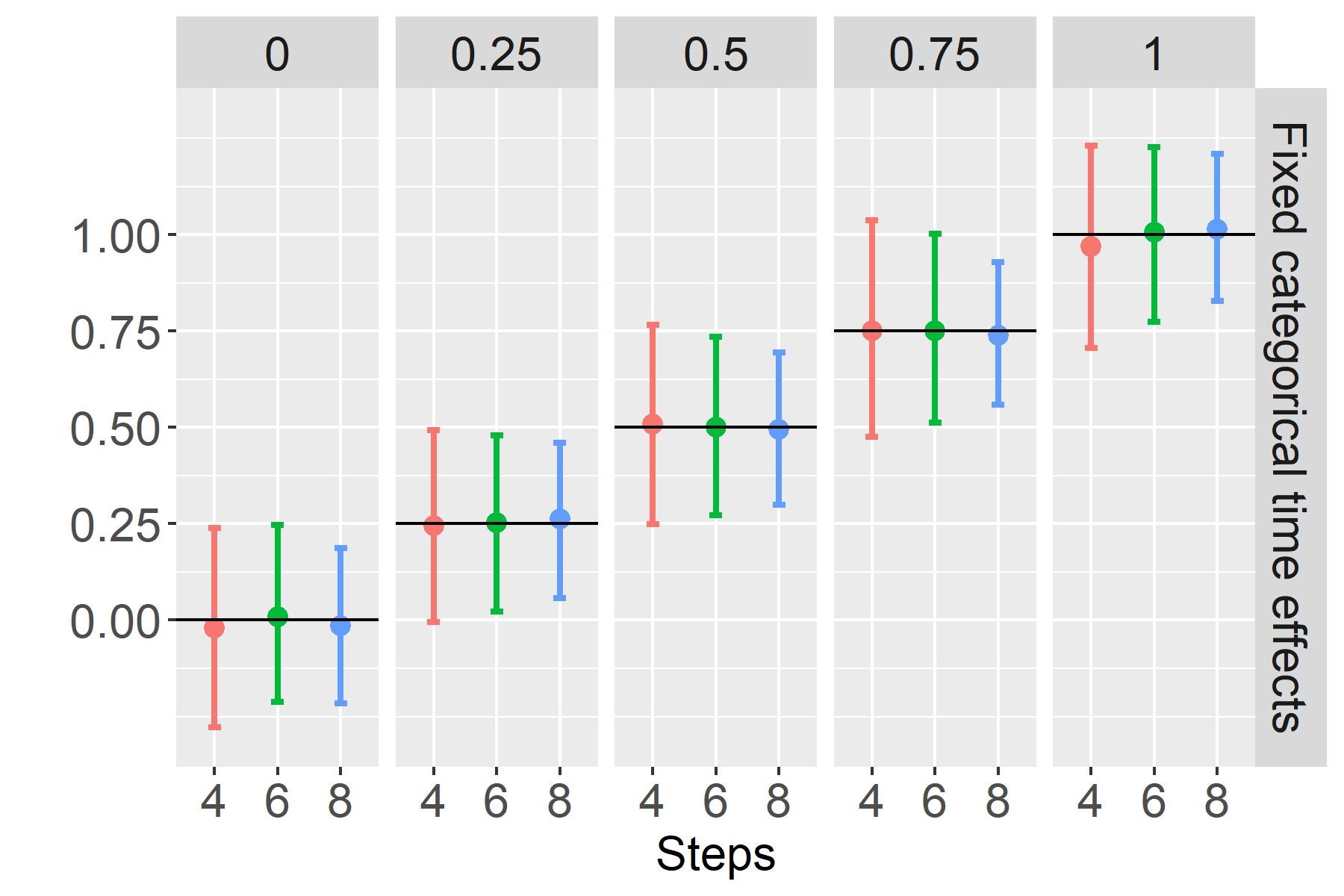}}
         \caption{Open cohort data}
         \label{fig:Estimate_linear_open}
     \end{subfigure}
        \caption{Mean and standard error of intervention effect estimates given by an LME model with fixed categorical time effects (model formulation~\ref{AnalysisModel_PeriodEff}) across 1000 simulations per parameter combination ($\theta$ = 0, 0.25, 0.5, 0.75, and 1; $J$ = 4, 6, and 8). Results are shown for different data scenarios: (a) closed cohort data where a continuous covariate influences the outcome linearly, (b) closed cohort data where a continuous covariate influences the outcome nonlinearly, (c) closed cohort data where a secular trend influences the outcome, and (d) open cohort data.}
        \label{fig:Estimate_ModelTimeEff_all}
\end{figure}

Figure~\ref{fig:Estimate_ModelTimeEff_all} shows that an analysis model with fixed time effects for every time period (model formulation~\ref{AnalysisModel_PeriodEff}) estimates the intervention effect $\theta$ unbiased (see also figure~\ref{fig:Bias_all} in the appendix) in different data scenarios:
\begin{itemize}
  \item[(a)] a continuous covariate, e.g., participants' age, influences the outcome linear,
  \item[(b)] a continuous covariate, e.g., participants' age, influences the outcome nonlinear,
  \item[(c)] a secular trend, e.g., due to COVID-19-related lockdowns, influences the outcome, and
  \item[(d)] open cohort data, i.e., 15\% of the participants are replaced from step to step.
\end{itemize}

\par
In data scenario (a), the linear predictor includes the intervention effect as well as the participants' age and the widowhood status. In data scenario (b), participants' age has a nonlinear influence on the outcome. Data scenario (c) differs from (a) as the outcome is additionally influenced by a secular trend. The same also applies to data scenario (d), whereby this is additionally an open cohort design, where, on average, the oldest 15\% per cluster and period leave the cohort and younger patients join. With an increasing number of steps and thus more outcome measurements per participant, the standard error of intervention effect estimates calculated for 1000 simulated datasets becomes smaller.

\FloatBarrier

\begin{figure}[!htb]\captionsetup[subfigure]{font=normalsize}
     \begin{subfigure}[t]{0.49\textwidth}
         \raisebox{-\height}{\includegraphics[width=\textwidth]{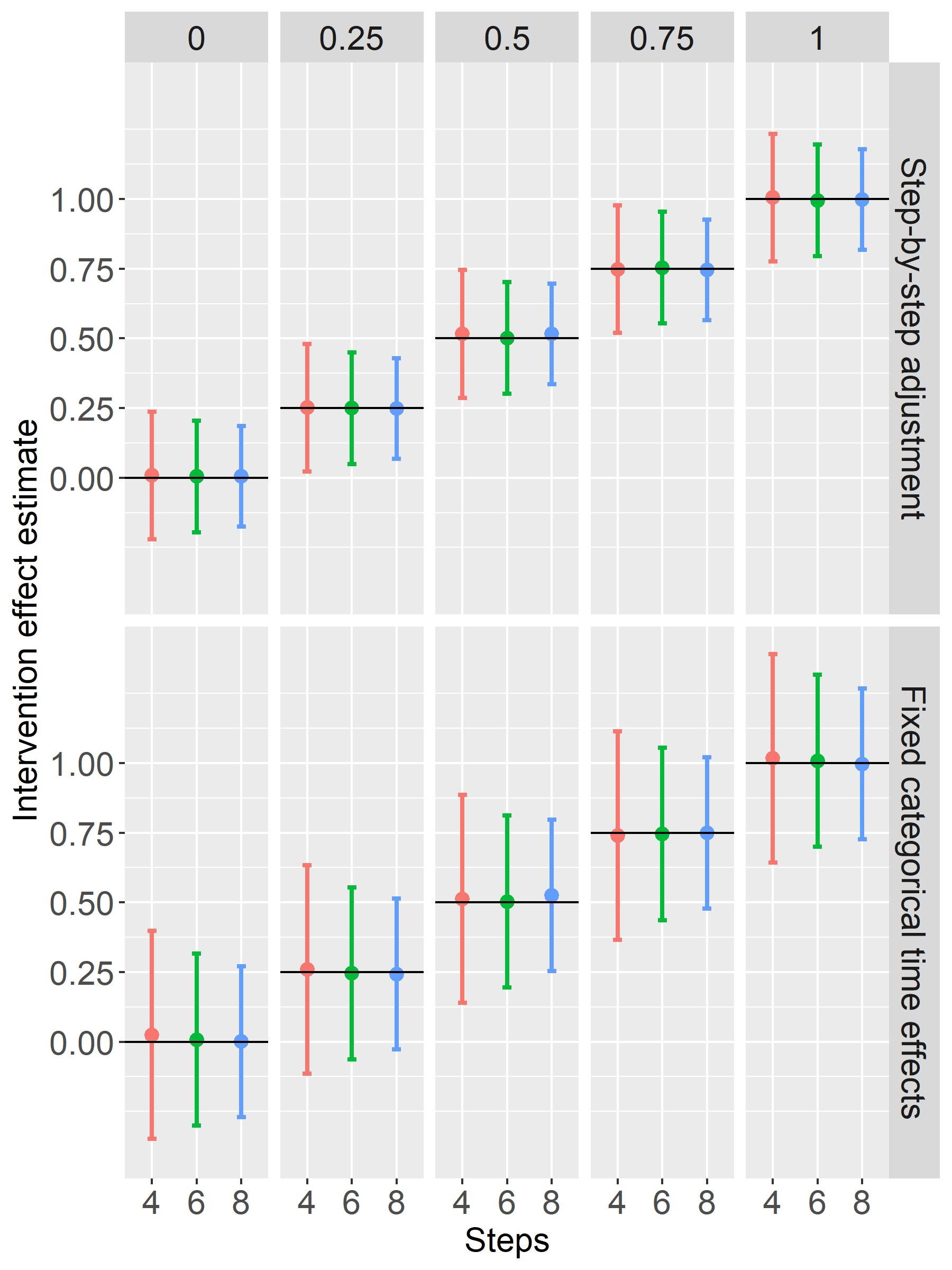}}
         \caption{Closed cohort data - Linear influence of continuous covariate}
         \label{fig:Estimate_linear_age_binary_StepAdj}
     \end{subfigure}
     \hfill
     \begin{subfigure}[t]{0.49\textwidth}
         \raisebox{-\height}{\includegraphics[width=\textwidth]{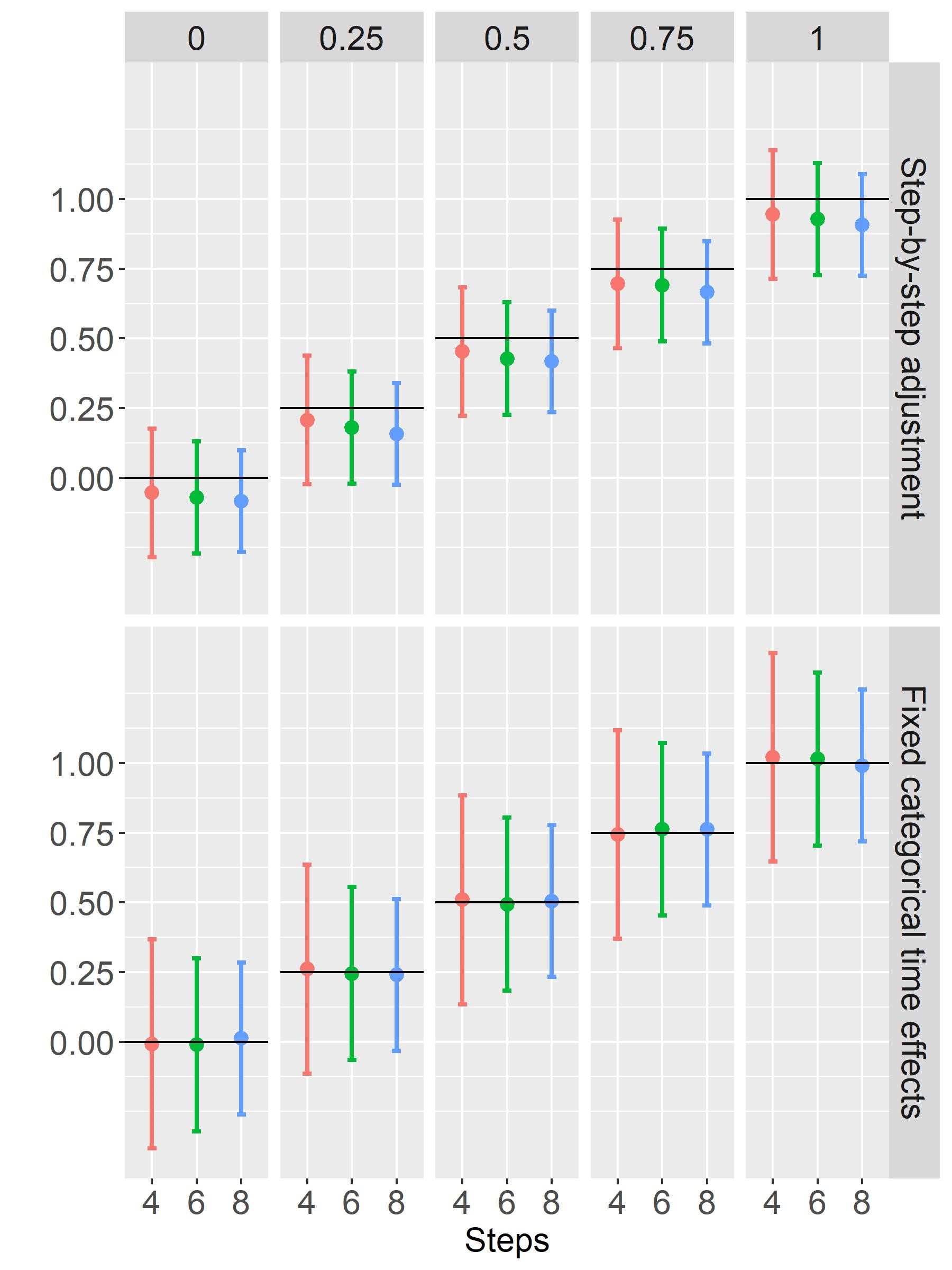}}
         \caption{Closed cohort data - Nonlinear influence of continuous covariate}
         \label{fig:Estimate_nonlinear_age_binary_StepAdj}
     \end{subfigure}.
        \caption{Mean and standard error of intervention effect estimates given by LME model formulation\ref{AnalysisModel_StepAdj} and ~\ref{AnalysisModel_PeriodEff} across 1000 simulations per parameter combination ($\theta$ = 0, 0.25, 0.5, 0.75, and 1; $J$ = 4, 6, and 8). Results are shown for two data scenarios: (a) closed cohort data where a continuous covariate influences the outcome linearly and (b) closed cohort data where a continuous covariate influences the outcome nonlinearly.}
        \label{fig:Estimate_StepAdj_all}
\end{figure}

Figure~\ref{fig:Estimate_linear_age_binary_StepAdj} shows that if continuous influence factors have a linear influence on the outcome, as in data scenario (a), an LME model with step-by-step adjustment of covariates (model formulation~\ref{AnalysisModel_StepAdj}) also provides unbiased intervention effect estimates. Moreover, this has the advantage that the power is greater (smaller IQR of intervention effect estimates; see also figure~\ref{fig:Power_all} in the appendix) than for the LME model with fixed categorical time effects (model formulation~\ref{AnalysisModel_PeriodEff}), since degrees of freedom are lost in estimating the time effect for each period. However, all confounders must be known and their functional influence on the outcome must be taken into account in the model formulation. If this is not the case, i.e., if the influence is formulated linearly but the covariate influences the outcome nonlinearly, Figure~\ref{fig:Estimate_nonlinear_age_binary_StepAdj} shows that the estimate of the intervention effect is biased.

\FloatBarrier

\begin{figure}[!htb]
 \centerline{\includegraphics[width=0.9\textwidth]{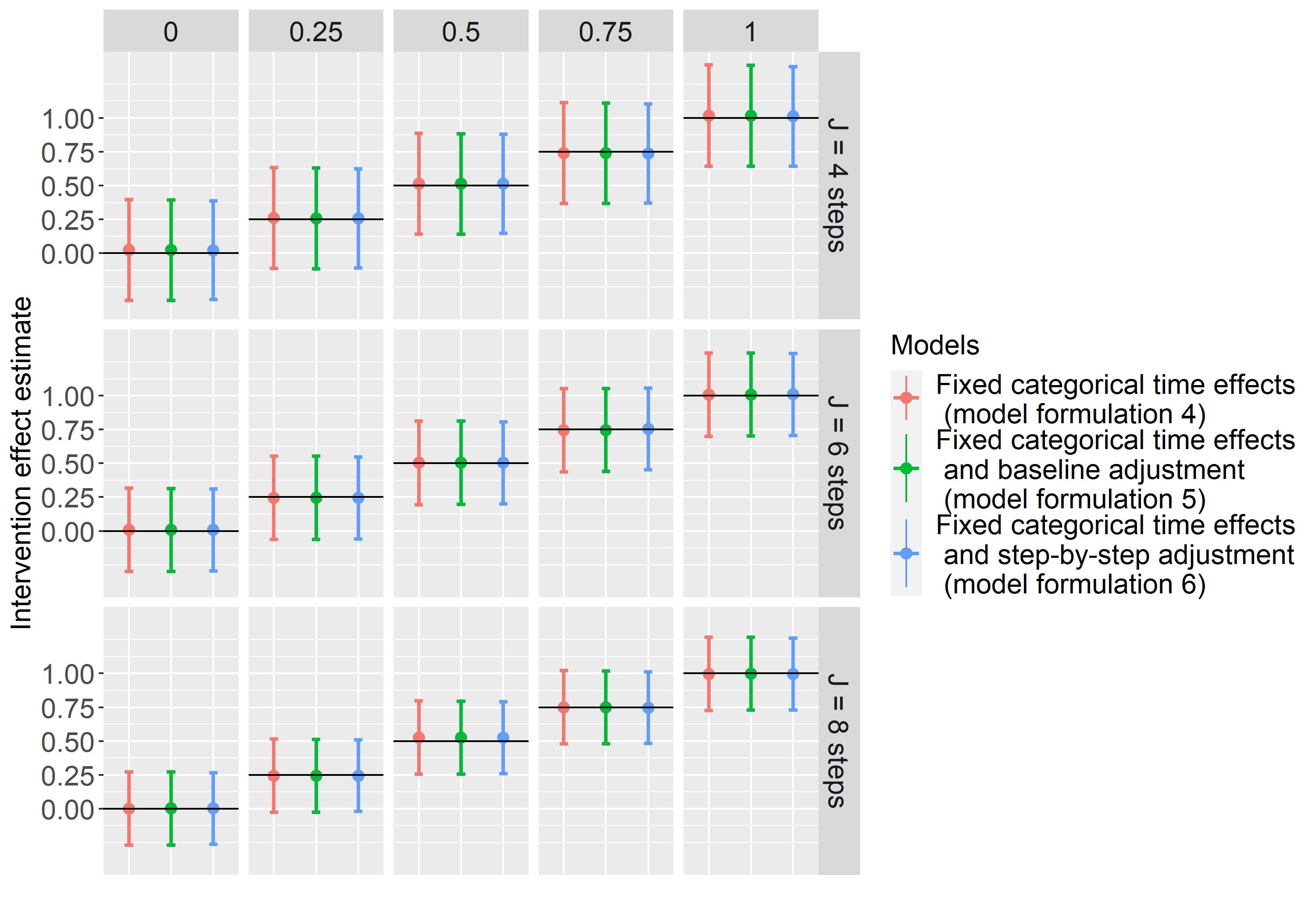}}
 \caption{Mean and standard error of intervention effect estimates given by LME model formulation~\ref{AnalysisModel_PeriodEff} (red), model formulation~\ref{AnalysisModel_PeriodEffBaselineAdj} (green), and model formulation~\ref{AnalysisModel_PeriodEffStepAdj} (blue) across 1000 simulations per parameter combination ($\theta$ = 0, 0.25, 0.5, 0.75, and 1; $J$ = 4, 6, and 8).}
        \label{fig:Estimate_TimeEffBaselineStep_all}
\end{figure}

Figure~\ref{fig:Estimate_TimeEffBaselineStep_all} shows that the combination of fixed categorical time effects and fixed effects for known time-varying covariates, adjusted for their baseline value or step-by-step, does not seem to increase the precision of the intervention effect estimators and thus does not lead to a power gain compared to a model with only fixed categorical time effects (see also figure~\ref{fig:Power_PeriodTimeEff_Compare3Models} in the appendix).

\FloatBarrier

\section{Discussion and conclusion}\label{chap::discussion}
So far the methodological focus of SW-CRTs has been on cross-sectional designs and guidance on the evaluation of SW-CRTs in cohort designs is much sparser. For this reason, previously published results on simulation studies \parencite{Thompson2017_B, Nickless2018, Rennert2021} investigated only the effect of time through secular trends which occur in both cross-sectional and cohort SW-CRTs. In the cohort design, however, in addition to external time effects internal time effect occur. Our simulation study is the first one investigating the influence of internal time effects in cohort SW-CRTs. It demonstrates that LME models with fixed categorical time effects additional to the fixed effect of intervention and two random effects used to account for the within-cluster and within-individual correlation (model formulation~\ref{AnalysisModel_PeriodEff}) seem to produce unbiased estimates of the participant specific intervention effect in SW-CRTs with closed and open cohort data even if time-varying confounders or their functional influence on outcome were unmeasured or unknown and if secular time trends occurred. But treating time as categorical (a separate parameter for each step) has a cost of wider confidence intervals compared to a model that considers all covariates adjusted step-by-step and according to their functional influence. Therefore, models with fewer parameters, and thus requiring fewer degrees of freedom to estimate parameters, have greater power to detect the same intervention effect. However, if the functional influence of the covariate on the outcome is misspecified, e.g., linear if it is nonlinear, this results in a biased estimate of the intervention effect. But avoiding biased intervention effect estimates should be a priority at any cost. A power gain by adding known confounders in the LME model formulation with categorical time effects cannot be found. The reason for this might be that the random patient effects already accounts for differences between the patients' characteristics at baseline, and the categorical time effects account not only for secular trends but also for changes in these characteristics over time.
\par
Due to the complexity of cohort SW-CRTs, we had to make some assumptions for our simulation. First, we assumed that in a closed cohort design each cluster contains the same number of participants, and second, we limited our simulation study to continuous outcomes. Third, we considered only models where the intervention effect is the same for all clusters and does not change over time, and where, in addition, the external time trend is the same for all clusters.
\par
A simulation study by Martin et al \parencite{Martin2019} provides evidence for cross-sectional SW-CRTs that unequal cluster sizes affect the precision of the intervention effect estimates and thus the power of an SW-CRT in the cross-sectional design, but not the bias. It is reasonable to assume that this is also true for cohort SW-CRTs. In addition, because recruitment of participants in closed cohorts occurs only at baseline, clusters are more likely to share a common size than in cross-sectional studies in which new patients are enrolled at each step.
\par
As has been done in other simulation studies \parencite{Thompson2017_B, Nickless2018}, we focus on one type of outcome. Consistent with the outcome under consideration in our motivating example, we have limited our simulation study to a continuous outcome. In research areas such as evidence-based medicine and health services research, where SW-CRTs are commonly used \parencite{Matthews2017}, patient-reported outcomes measures are gaining popularity for comparing patient groups \parencite{Kohlmann2010}. Typically differences in group means are carried out using sum-score-based approaches, which are mostly continuous measurements.
\par
A variety of simulation studies investigated bias from misspecified mixed-effect models in cross-sectional SW-CRTs \parencite{Thompson2017_B, Voldal2022, Kenny2022}. Model misspecification could result from different influences of nonconstant intervention effects that were not adequately accounted for. For simplicity, our simulation study focused on constant intervention effects over time and across clusters. However, it seems reasonable to assume that most of the results of the other simulation studies can be generalized to nonconstant intervention effects in cohort SW-CRTs.
\par
From our simulation study, we can conclude that, in addition to the fixed intervention effect, fixed categorical time effects and random cluster and participant effects seem to be sufficient to compute unbiased estimators for the intervention effect, regardless of whether the study cohort characteristics change over time (regarding age, marital status,...). This is a great advantage, as one does not have to think about possible confounders and their functional influence on the outcome. Whilst further research is needed to explore the potential for bias in different data settings in cohort SW-CRTs, we contribute to the literature's attention the particularities of cohort SW-CRTs compared to SW-CRTs in a cross-sectional design. To provide trialists more guidance on evaluating cohort SW-CRTs, the results of our simulation study on handling internal time effects should be included in a next version of the CONSORT statement on reporting SW-CRTs.


\section*{Acknowledgments}
We would like to thank Matthew Cooper for the professional language review and Tobias Schipp for his contributions to the simulation code. Furthermore, we want to thank Robin Denz, Marianne Tokic, and Dr. rer. nat. Hans Diebner for their helpful comments and suggestions in our biweekly PhD Colloquium. This research was conducted as part of the PhD thesis of Jale Basten at TU Dortmund University.

\newpage

\printbibliography

\newpage

\appendix

\section{Additional Plots of Simulation Results}

\begin{figure}[!htb]\captionsetup[subfigure]{font=normalsize}
     \begin{subfigure}[t]{0.49\textwidth}
         \raisebox{-\height}{\includegraphics[width=\textwidth]{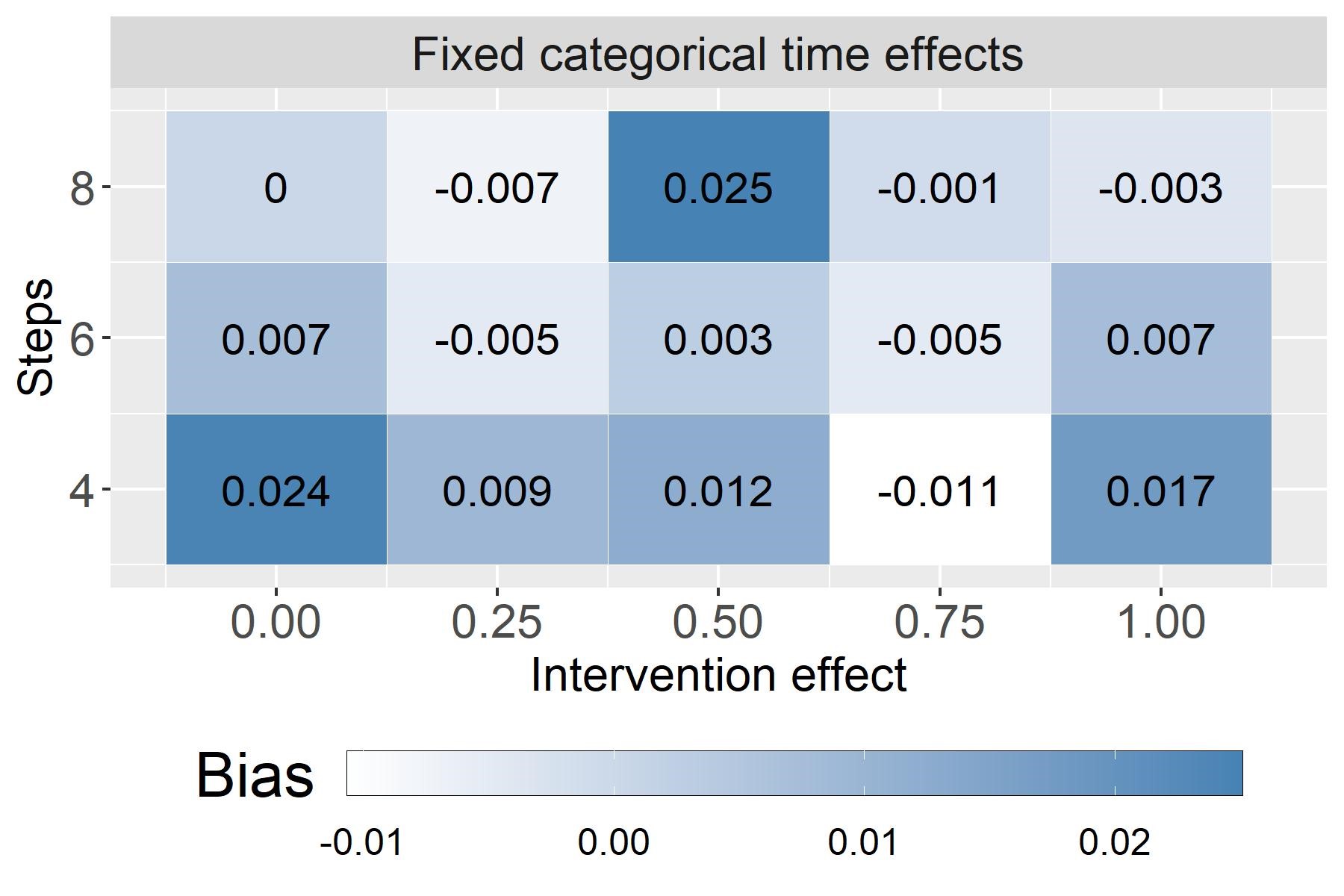}}
         \caption{Closed cohort data - Linear influence of time-varying continuous covariate}
         \label{fig:Bias_linear_age_binary}
     \end{subfigure}
     \hfill
     \begin{subfigure}[t]{0.49\textwidth}
         \raisebox{-\height}{\includegraphics[width=\textwidth]{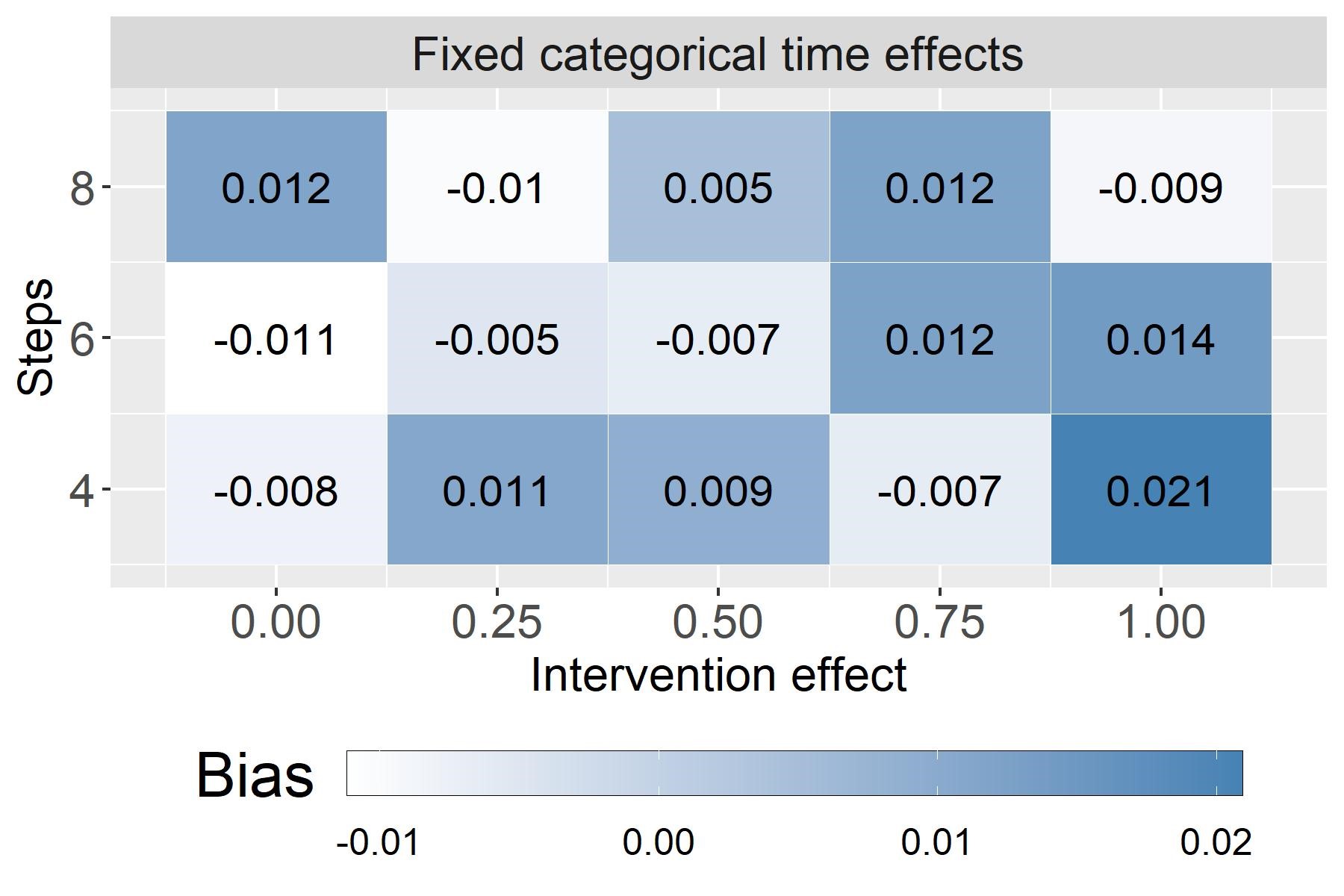}}
         \caption{Closed cohort data - Nonlinear influence of time-varying continuous covariate}
         \label{fig:Bias_nonlinear_age_binary}
     \end{subfigure}
     \begin{subfigure}[t]{0.49\textwidth}
         \centering
         \raisebox{-\height}{\includegraphics[width=\textwidth]{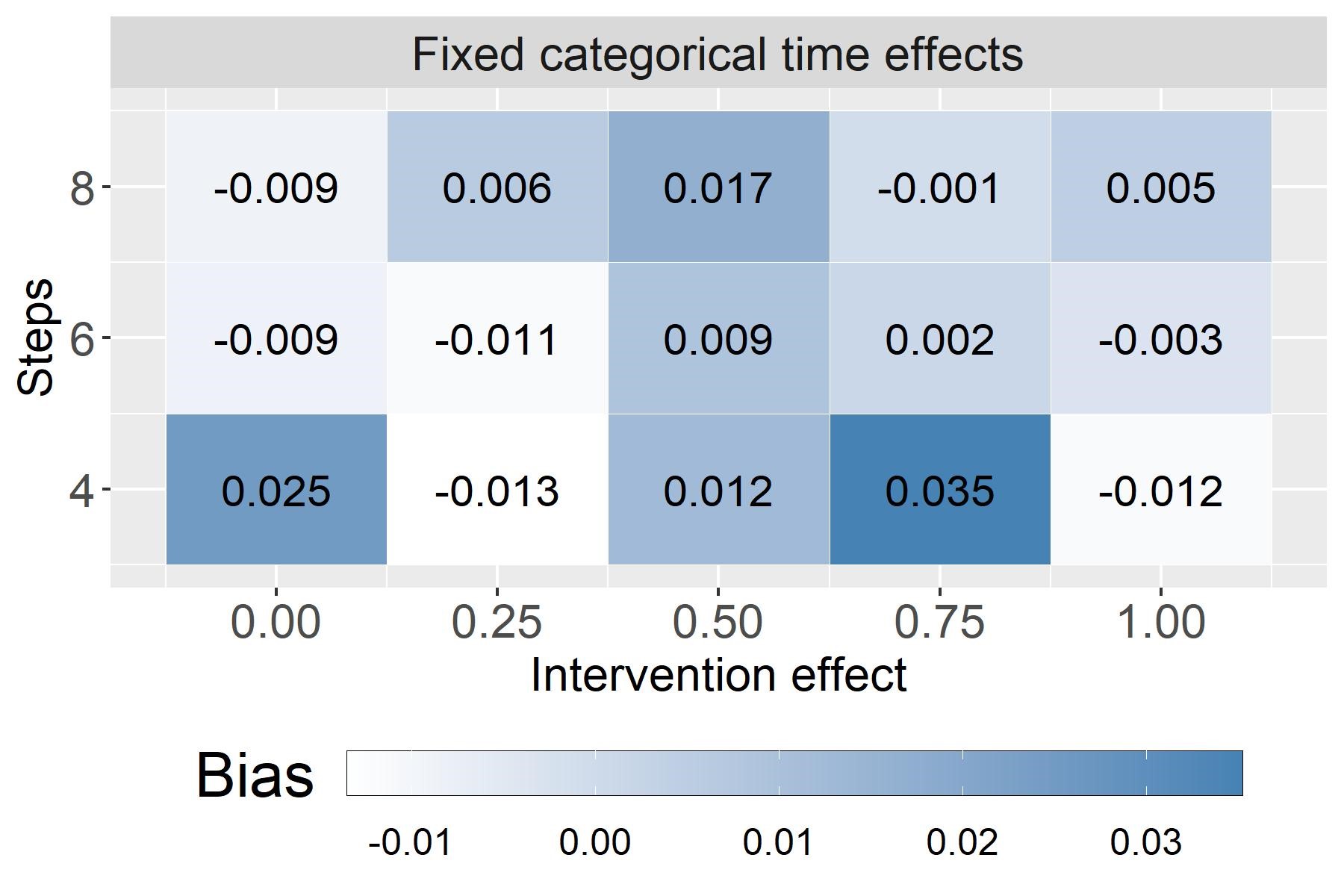}}
         \caption{Closed cohort data - Secular trend}
         \label{fig:Bias_linear}
     \end{subfigure}
     \hfill
     \begin{subfigure}[t]{0.49\textwidth}
         \centering
         \raisebox{-\height}{\includegraphics[width=\textwidth]{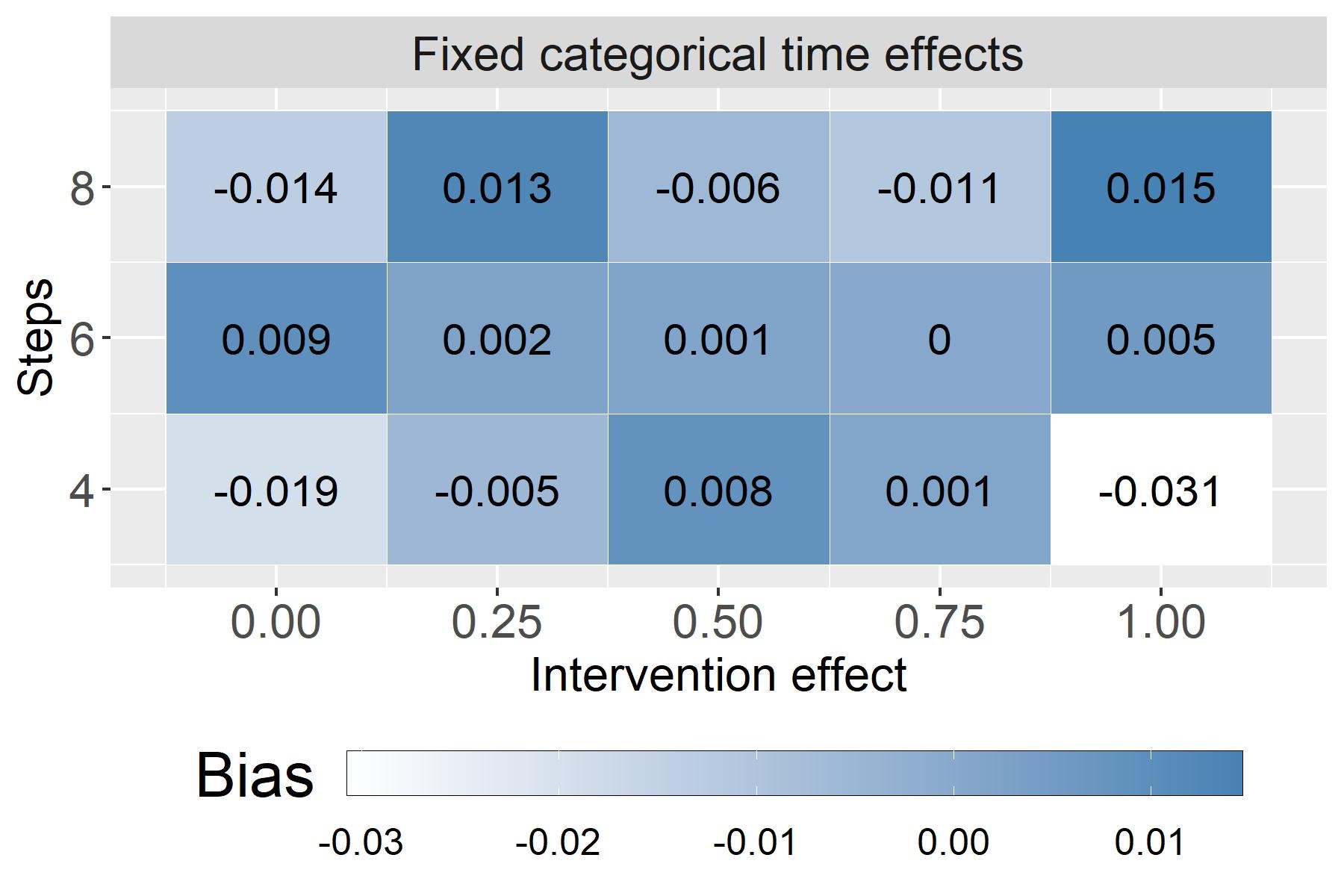}}
         \caption{Open cohort data}
         \label{fig:Bias_linear_open}
     \end{subfigure}
     \caption{Bias of intervention effect estimates given by LME model formulation~\ref{AnalysisModel_PeriodEff} across 1000 simulations per parameter combination ($\theta$ = 0, 0.25, 0.5, 0.75, and 1; $J$ = 4, 6, and 8). Results are shown for different data scenarios: (a) closed cohort data where a time-varying continuous covariate influences the outcome linear, (b) closed cohort data where a time-varying continuous covariate influences the outcome nonlinear, (c) closed cohort data where a secular trend influences the outcome, and (d) open cohort data.
     \label{fig:Bias_all}}
\end{figure}
\FloatBarrier

\begin{figure}[!htb]
 \centerline{\includegraphics[width=0.8\textwidth]{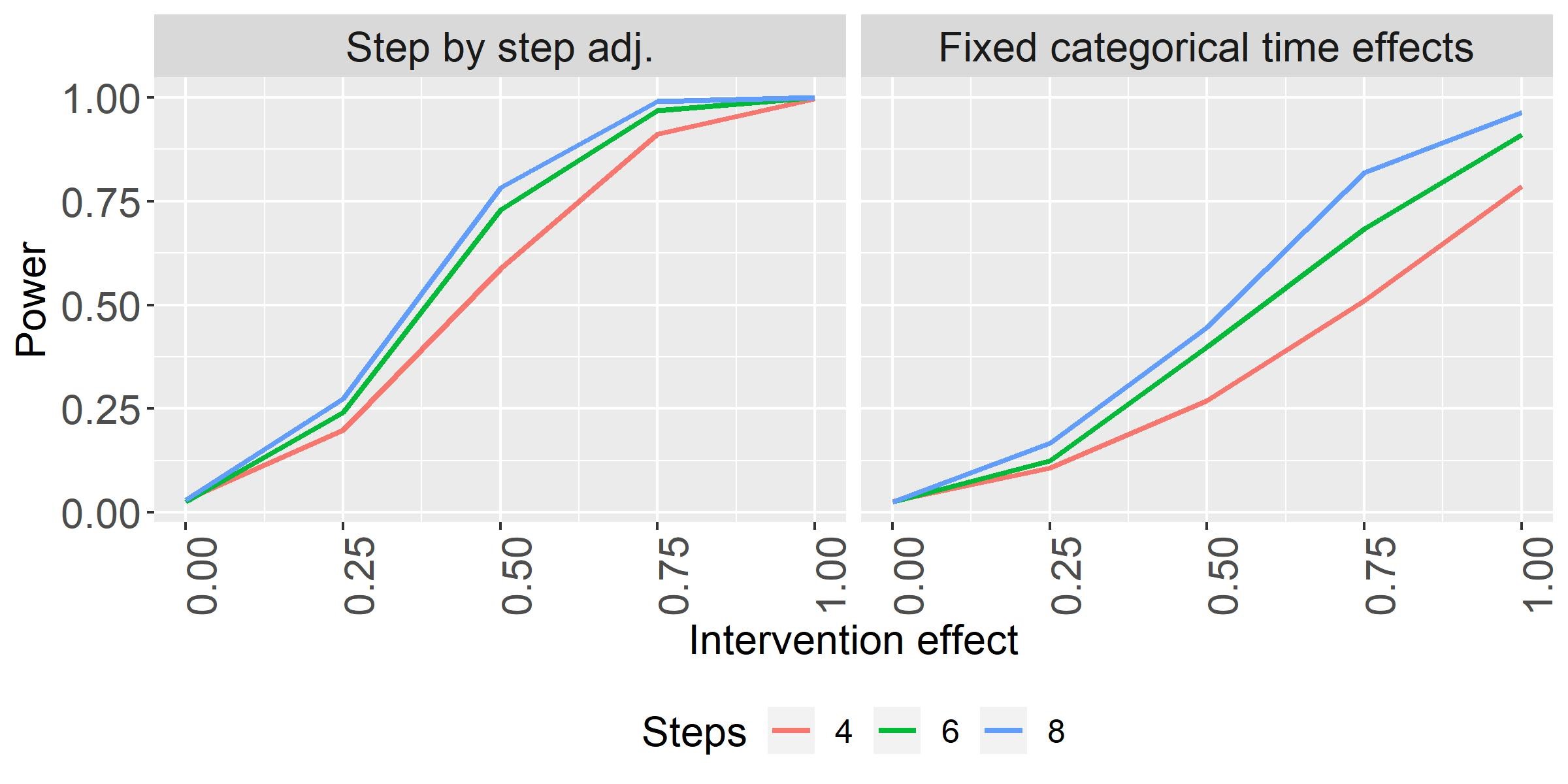}}
 \caption{Power estimating the intervention effect by LME model formulation~\ref{AnalysisModel_StepAdj} and~\ref{AnalysisModel_PeriodEff} across 1000 simulations per parameter combination ($\theta$ = 0, 0.25, 0.5, 0.75, and 1; $J$ = 4, 6, and 8).\label{fig:Power_all}}
\end{figure}
\FloatBarrier

\begin{figure}[!htb]\captionsetup[subfigure]{font=normalsize}
     \begin{subfigure}[t]{\textwidth}
         \raisebox{-\height}{\includegraphics[width=\textwidth]{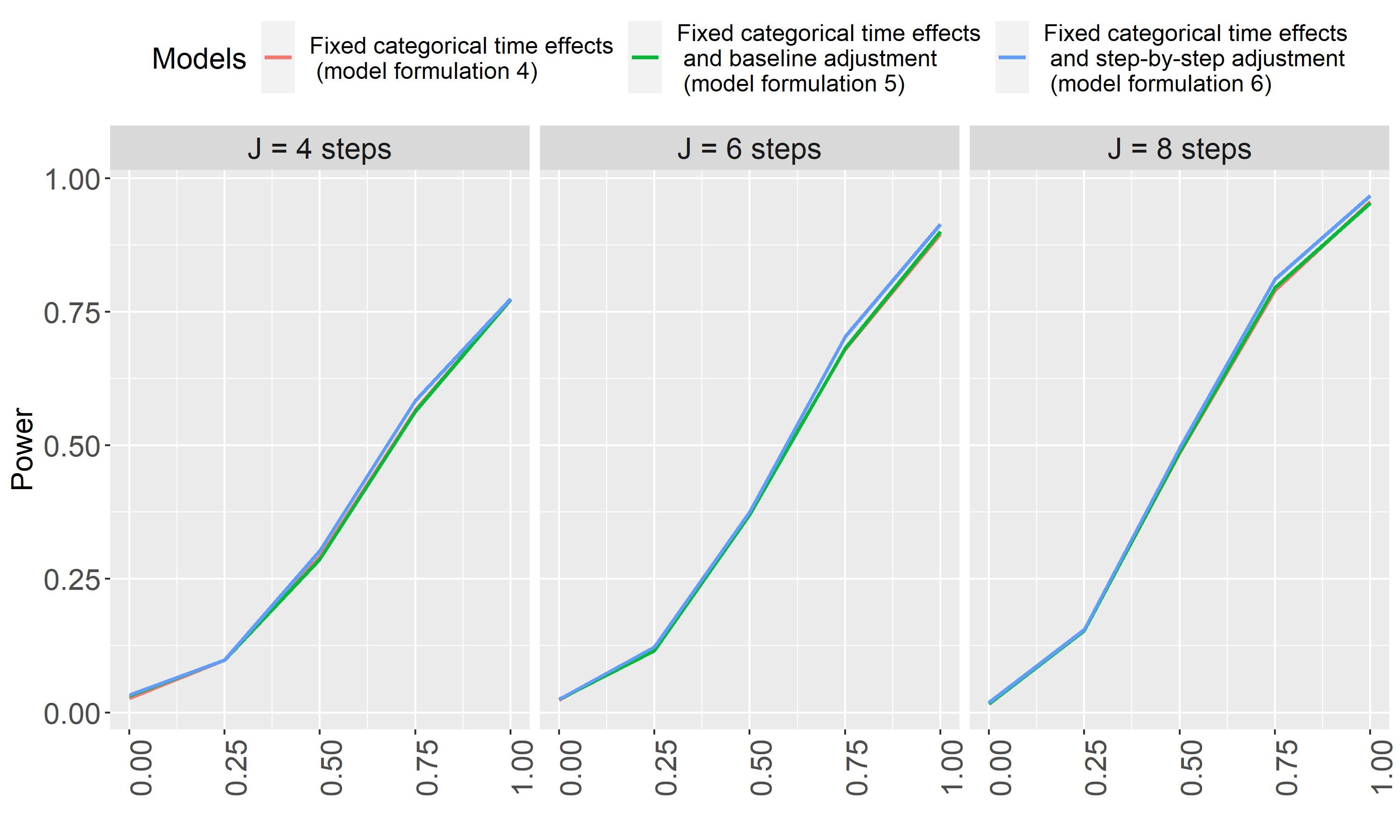}}
         \caption{Linear influence}
         \label{fig:Power_PeriodTimeEff_Compare3Models_linear}
     \end{subfigure}
     \begin{subfigure}[t]{\textwidth}
         \centering
         \raisebox{-\height}{\includegraphics[width=\textwidth]{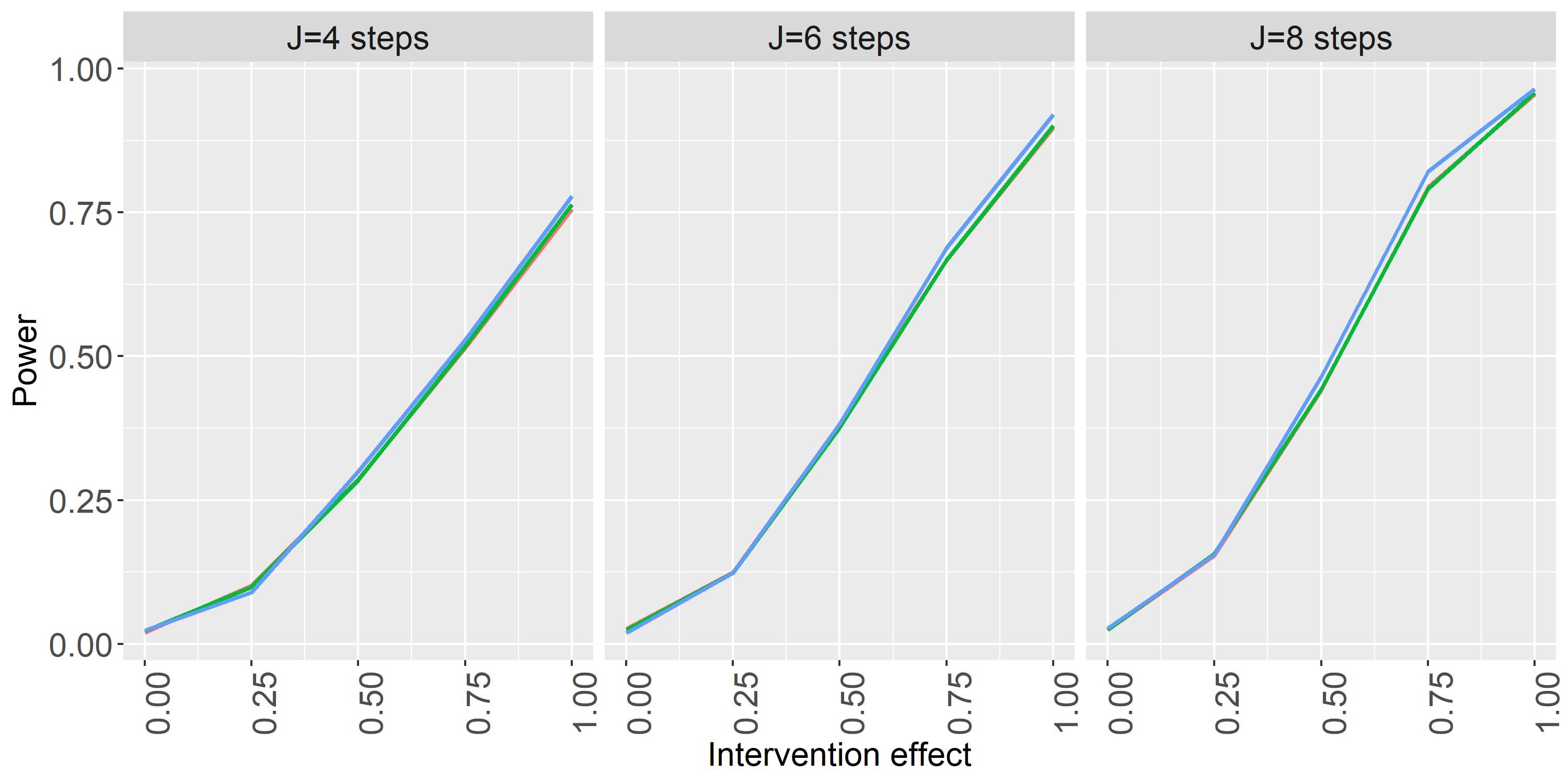}}
         \caption{Nonlinear influence}
         \label{fig:Power_PeriodTimeEff_Compare3Models_nonlinear}
     \end{subfigure}
     \caption{Power estimating the intervention effect by LME model formulations~\ref{AnalysisModel_PeriodEff},~\ref{AnalysisModel_PeriodEffBaselineAdj}, and~\ref{AnalysisModel_PeriodEffStepAdj} across 1000 simulations per parameter combination ($\theta$ = 0, 0.25, 0.5, 0.75, and 1; $J$ = 4, 6, and 8).\label{fig:Power_PeriodTimeEff_Compare3Models}}
\end{figure}
\FloatBarrier

\end{document}